\documentclass[useAMS,usenatbib]{mn2e}
\usepackage{graphicx}
\usepackage{aas_macros}
\usepackage{cite}
\usepackage{array}
\title[Wolf-Rayet Ejecta Nebulae]{A search for Ejecta Nebulae around Wolf-Rayet Stars using the SHS H$\alpha$ survey} %have to be the same?

\author[D. J. Stock \& M. J. Barlow]{D.J. Stock$^{1}$\thanks{E-mail:
dstock@star.ucl.ac.uk (DS)} \& M.J. Barlow$^{1}$\\
$^{1}$Department of Physics \& Astronomy, University College London, Gower Street, London, WC1E 6BT, UK\\
}
\begin{document}

\date{\today}

\date{Accepted 2010 June 2.  Received 2010 May 30; in original form 2010 April 9}
\pagerange{\pageref{firstpage}--\pageref{lastpage}} \pubyear{2009}

\maketitle

\label{firstpage}

\begin{abstract}

Recent large scale Galactic Plane H$\alpha$ surveys allow a re-examination of the environs of Wolf-Rayet (WR) stars for the presence of a circumstellar
nebula. Using the morphologies of WR nebulae known to be composed of stellar ejecta as a guide, we constructed ejecta nebula criteria similar to those of \citet{1991IAUS..143..349C} and searched for likely WR ejecta nebula in the SHS H$\alpha$ survey. A new Wolf-Rayet ejecta nebula around WR 8 is found and its morphology discussed. The fraction of WR stars with ejecta type nebulae is roughly consistent between the Milky Way (MW) and LMC at around 5-6\%, with the MW sample dominated by nitrogen rich WR central stars (WN type) and the LMC stars having a higher proportion of carbon rich WR central stars (WC type). We compare our results with those of previous surveys, including those of \citet{M97} and \citet{MC93}, and find broad consistency. We investigate several trends in the sample: most of the clear examples of ejecta nebulae have WNh central stars; and very few ejecta nebulae have binary central stars. Finally, the possibly unique evolutionary status of the nebula around the binary star WR 71 is explored.

\end{abstract}

\begin{keywords}
Wolf-Rayet, Chemical Evolution, Stellar Evolution, Ring Nebula, Massive Stars
\end{keywords}

\section{Introduction}

Since the suggestion that some Wolf-Rayet stars could be creating nebulae via mass loss \citep{JH65},  efforts to detect more examples have been ongoing. The first attempt to morphologically categorise nebulae presumed to have been created by the influence of WR stars was performed by \citet{C81}, who devised three broad categories for possible nebulae: W, R \& E. 

W type nebulae are those which are assumed to be $``$Wind-Blown Bubbles$"$. These objects were inferred to have been created when material ejected from the star interacted with the surrounding ISM. R type nebulae were postulated to be those in which the strong radiation field from the host star excites the surrounding ISM. These are only visible when the surrounding ISM is of sufficient density to produce detectable emission. 

Our interest lies with the E type nebulae, which were defined to be those which were likely to contain processed ejecta from the progenitor star. These were suggested by Chu to have been formed by a violent mass loss episode recently in the star's history, which may not have been isotropic or homogeneous. The nebulosity can therefore be very clumped and irregular. The lifetime of E type nebulae should be much lower than for the other two types, since as the ejected mass expands as it moves away from the star, its surface brightness should diminish very quickly. It should be noted that actual Wolf-Rayet nebulae can (and do) display any combination of the above traits. \citet{1991IAUS..143..349C} modified the scheme to refine the definition of E type nebulae, splitting the category into Stellar Ejecta nebulae and Bubble/Ejecta (W/E) nebulae. The former covered pure E type nebulae as defined above, the latter introduced to cover the case of ejecta shells having merged with the swept up shell.

The \citet{C81,1991IAUS..143..349C} criteria were used for the southern galactic plane surveys of \citet{M94a, M94b} and \citet{M97} but not for the northern survey of \citet{MC93} which described the probability of a ``ring" nebula being present. The survey of the environs of all Magellanic Cloud WR nebulae by \citet{D94} also ignored the Chu categorisation system and commented only on the presence of a ring or whether the star was in a superbubble. 

In this work we shall consider only morphological information in our classifications of new nebulae and will not delve into the physical details of their formation and evolution. Theory and models of the formation of nebulae around massive stars were presented by \citet{1991IAUS..143..349C, 2003IAUS..212..585C}.

Recent publicly available H$\alpha$ surveys \citep{D05,P05} allow re-inspection of the environs of all known WR stars with a view to identifying new E type nebulosities which can provide constraints on the nucleosynthetic effects of WR stars.

\section{Morphologies of Spectroscopically confirmed WR Ejecta nebulae}

\begin{figure*}
	\centering
	\includegraphics[width=170mm]{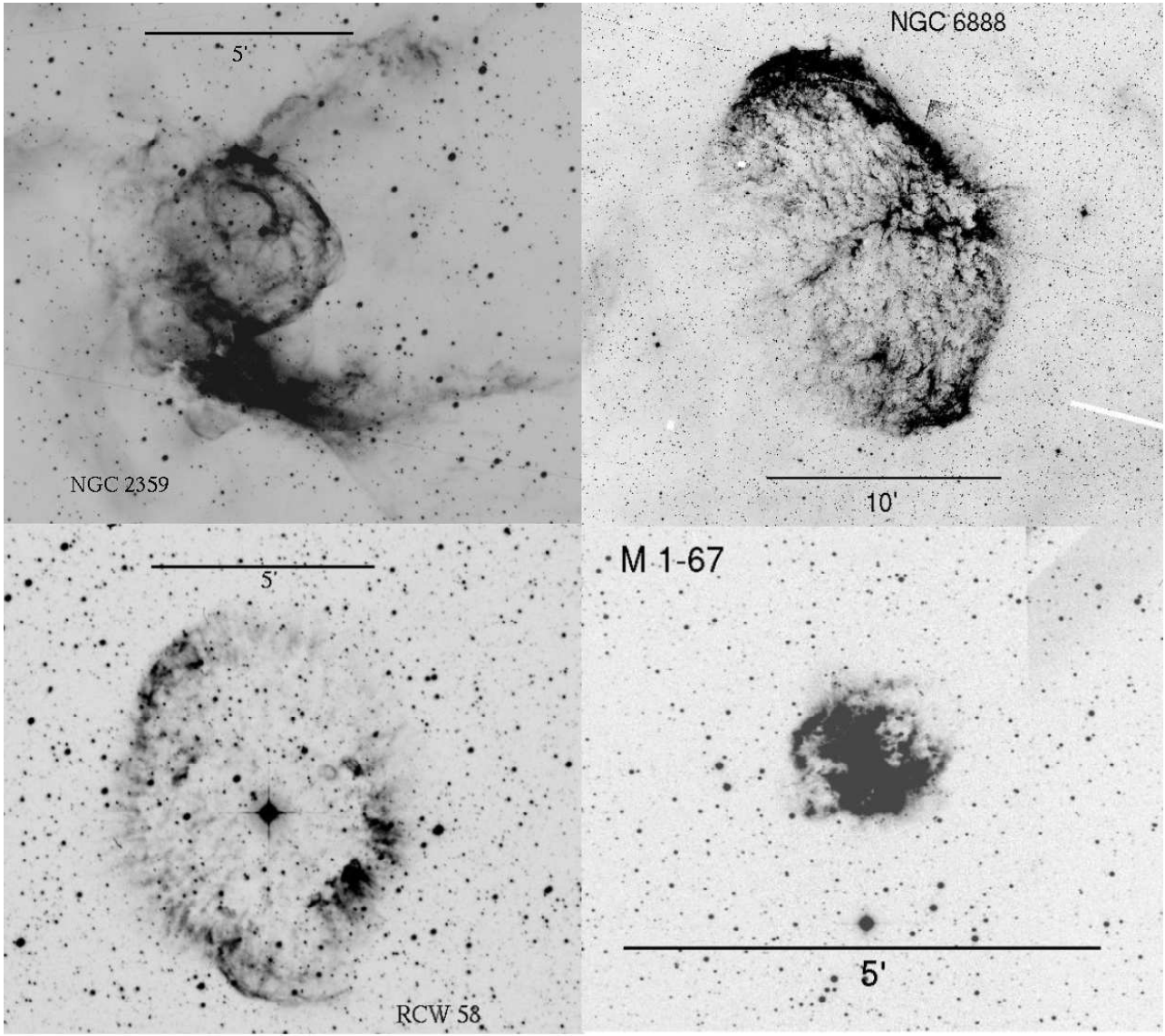}
	\caption{Morphologies of WR Ejecta nebulae spectroscopically confirmed to have a chemically processed component. Clockwise from top left: SHS imagery of NGC 2359 (WR 7,WN4); IPHAS montage of NGC 6888 (WR 136, WN6h), kindly supplied by Nick Wright, IfA Harvard; IPHAS image of M1-67 (WR 124, WN8h); SHS imagery of RCW 58 (WR 40, WN8h). North is up and east to the left in all images. }
	\label{confirmed}
\end{figure*}

The W, R and E categories of \citet{C81} and \citet{1991IAUS..143..349C} were tested over the following decades as spectroscopic data for the different subtypes of WR nebulae began to appear.

\begin{table*}
	\centering
	\caption{Nebulae Spectroscopically Confirmed to contain Nucleosynthetic products  }
	\label{SpecNeb}
	\begin{tabular}{ccccrrp{3cm}}
		\hline
		WR$^a$ & Spectral Type$^b$ & Nebular Name & Classification$^c$ & Reference & Enrichment &  Reference   \\
		\hline
		6   & WN4  & S308      & W   & \citet{1991IAUS..143..349C} & N and He enriched       & \citet{1992AA...259..629E} \\
		7   & WN4  & NGC 2359  & W   & \citet{1991IAUS..143..349C} & enriched He knots       & \citet{E90} \\
		16  & WN8h & Anon      & W/E & \citet{M97}                 & enriched N throughout   & \citet{M99} \\
		40  & WN8h & RCW 58    & W/E & \citet{1991IAUS..143..349C} & N and He enriched       & \citet{K84} \\
		75  & WN6  & RCW 104   & W/E & \citet{1991IAUS..143..349C} & N enrichment            & \citet{1992AA...259..629E} \\
		102 & WO2  & G 2.4+1.4 & W   & \citet{1991IAUS..143..349C} & enriched He             & \citet{E91} \\
		124 & WN8h & M 1-67    & E   & \citet{1991IAUS..143..349C} & N overabundance         & \citet{E91} \\
		136 & WN6h & NGC 6888  & W/E & \citet{1991IAUS..143..349C} & Strong N, He enrichment & \citet{E92-6888}, \citet{M00-6888} \\

	\end{tabular}
	
	\textit{a}: Catalogue numbers from \citet{H00}.\\
	\textit{b}: Spectral types from \citet{H00}.	
	\textit{c}: Using the \citet{1991IAUS..143..349C} scheme
	
\end{table*}

For the two nebulae which \citet{C81} regarded as E type (M 1-67 and RCW 58) the nebulosities were found to be enriched relative to the ISM in nitrogen and helium but depleted in oxygen \citep{K84,E91}. The anonymous nebula surrounding WR 16 was also shown to be comprised of material with a very similar abundance pattern to those of M1-67 and RCW 58 \citep{M99}. The detection of processed material in these nebulae was a major success for the categorisation scheme, as this showed that it is possible to infer likely patterns in the chemical composition of a WR nebula by studying its morphology.

However, material displaying the same patterns of enrichment was also detected in NGC 6888 - a nebula Chu had initially classified as W type \citep{E92-6888,M00-6888}. This showed clearly that the lines between the initial Chu classes can be blurred, indeed NGC 6888 - see Figure \ref{confirmed}, top right - appears to be a mixture of different kinds of nebulosity. The edge looks like a wind-blown shell, whilst there is evidence of flocculent nebulosity in the central regions, suggesting ejected material. NGC 6888 was later re-classified as a W/E nebulae by \citet{1991IAUS..143..349C}.

\citet{GC93} noted that the nebulosity around BAT99-16 in the LMC displays the same abundance pattern as RCW 58, M 1-67, WR 16 \& NGC 6888.

The results of \citet{E90} \& \citet{E91} also suggested another class of WR nebula, those containing helium overabundances without any enrichment in heavier elements. Two nebulae, NGC 2359 \citep{E90} and G 2.4+1.4 \citep{E91}, which present this abundance pattern display striking similarities, but appear to have been created in different ways. G 2.4+1.4 was initially classed as a supernova remnant \citep{DP90}, but \citet{1991IAUS..143..349C} categorised it as W type. G 2.4+1.4 was later found to have a mass of 4200 $M_{\sun}$ \citep{GL02}  which implies that the nebulosity is primarily swept up, agreeing with the \citet{1991IAUS..143..349C} categorisation. NGC 2359 is a very clear example of a wind-blown bubble. The ionized mass of NGC 2359 has been estimated to be of around $70 M_{\sun}$ and the mass of the whole complex over $1000 M_{\sun}$ \citep{CG99}, which clearly precludes a pure ejecta origin. However, both nebulae display helium enrichment - showing again that the lines between Chu's classes can be blurred. 
    
The above spectroscopic results lead to the conclusion that the morphological criteria for ejecta nebulae presented by \citet{C81} may be too stringent, a problem addressed by the introduction of the Bubble/Ejecta (W/E) class \citep{1991IAUS..143..349C}. Wind-blown bubbles can also contain ejecta in their filamentary nebulosities (e.g. NGC 6888). 

The above spectroscopic information suggests the following, revised criteria: ejecta nebula candidates must have either a highly flocculent structure, as in Chu's scheme or, alternatively, possess flocculent structure within their wind-blown shells, similar to that shown by NGC 2359, NGC 6888 \& NGC 3199. 

Radiatively excited nebulae, (R type) are more difficult to classify under this scheme as they tend to possess highly irregular morphologies which are dependent not on stellar outflows but on the density distribution of material in the local ISM.

\section{Expected Wolf-Rayet Nebular Size Scales}

The angular scales over which Wolf-Rayet Nebulae have previously been detected span more than an order of magnitude, with the smallest being of the order of arcseconds in diameter (BAT99-2, LMC, \citet{D94}) and the largest were claimed to have an angular radius of a degree or more e.g. $\theta$ Muscae \citep{G98}. However the nebulosity surrounding $\theta$ Muscae has been shown not to be related to the star \citep{2010MNRAS.401.1760S} .

An upper limit on the angular radius of a nebula composed of ejecta from the WR or immediate pre-WR phase is given by: $\theta=\frac{\tau_{WR} \times V_{exp}}{D}$ where $V_{exp}$ is the expansion velocity of the  nebula, $D$ is the distance to the nebula and $\tau_{WR}$ is the time spent in the WR phase $\sim 3 \times 10^5 $ years \citep{C07}.  We have chosen this upper bound as a representative figure for the largest wind blown nebulae, as it is suspected that ejecta can expand much faster in the evacuated volume within a wind-blown bubble, allowing ejecta nebulae to possibly take on much larger sizes. $V_{exp}$ is a typical expansion velocity of a WR nebula: $\sim 25-50$ km/s \citep{C82A, C83} we assume this figure is typical for expanding shells around galactic WR stars. 

For a WR star at 1 kpc this predicts a maximum angular diameter of $\sim$ 45 arcmin. We can also estimate the distance threshhold below which 30 arcmin extractions from the SHS could be insufficient to encompass a WR nebula. Rearranging for distance and using $\theta = 0.5^{\circ}$, this yields a distance of $\sim$ 1.5 kpc.

\section{SHS H$\alpha$ imagery}

The AAO/UKST Southern H$\alpha$ Survey (SHS) took place during the early 2000's and covered the entire southern galactic plane \citep{P05}. The survey was one of the last examples of the use of photographic imaging in astronomy and demonstrates neatly the advantages and disadvantages of this approach. The survey was performed using the 1.2m UK Schmidt Telescope (UKST) at the Anglo-Australian Observatory (AAO), on 5 degree square photographic plates using three hour exposures. Each plate was subsequently transported to the Royal Observatory Edinburgh (ROE) and digitised using the SuperCOSMOS plate scanning machine at a resolution of 0.67 arcseconds/pixel. The data were then montaged and made publicly available via the internet. 

The advantages of photographic plates are twofold, firstly a large amount of data can be captured on each 25 square degree exposure, with comparatively little effort needed to extract images. Secondly the photographic emulsion was tuned such that the exposures were sensitive to very faint structure. The obvious disadvantage to utilising these data is the difficulty of flux calibration, since photographic emulsion is not a linear detector and intensity calibration can be very troublesome.

The SHS also covered the Magellanic Clouds and while these data were digitised by the SuperCOSMOS project they were never made publicly available. They were kindly made available to us by the Wide-Field Astronomy Unit (WFAU) at the ROE-IfA (Royal Observatory Edinburgh - Institute for Astronomy) .

The SuperCOSMOS interface to the SHS imagery allows extractions of up to 900 square arcminutes, in order to inspect WR nebulae larger than this the ``Montage" package, produced by NASA's IPAC (Infrared Processing and Analysis Center), was utilized to combine SHS extractions.

Northen galactic plane WR stars were examined using IPHAS survey data \citep{D05}. However as no disagreement was found between IPHAS imagery and the survey of \citet{MC93} no further discussion will be presented.   

\section{Results \& Comparisons with previous work}
   
\subsection{Results}

Each southern star in the Sixth catalogue of Galactic Wolf-Rayet Stars \citep{H00} and its annex \citep{2006AA...458..453V} was inspected visually using SHS imaging data as described above. The results are summarised in Table \ref{ProgTable}. Each WR star in the LMC catalogue of \citet{B99} (hereafter BAT99) was also visually inspected and resolvable nebulae categorised in the same way. SMC WR stars were also investigated but none appeared to possess resolvable ejecta nebulae. In addition, the positions of several WR ring nebulae that were discovered by \citet{2010AJ....139.2330W} at mid-infrared wavelengths were inspected - yielding one candidate ejecta nebula.                                 

For the LMC it is difficult to identify WR nebulae whose absolute sizes are as small as some Galactic WR nebulae such as M1-67, since their angular sizes would have been smaller than the overexposed PSF of the host star in SHS imagery. Conversely, if we place the largest LMC WR nebula: Anon (HD 32402) at 1 kpc, it would have an angular size of $1.5^{\circ}$ which would not have been discernable to our survey. However very few ($< 10$) WR stars are closer than 1 kpc. This effect means that we cannot see details on the same scales as their galactic counterparts; we can only determine likely candidates for further study.

\subsection{Comparison with previous work}

A comparison of our results with those presented by \citet{1991IAUS..143..349C} and by \citet{M97} is presented in Table \ref{Agreement}. 

\begin{table}
	\centering
	\caption{WR Nebular Classifications}
	\label{Agreement}
	\begin{tabular}{cp{1.2cm}p{1.2cm}p{1.2cm}p{1.2cm}}
		\hline
		Cat. No. & Spectral Type$^a$ & \citet{1991IAUS..143..349C}$^b$ & \citet{M97}$^b$ & This Work \\
		\hline                                                         
		WR 6   & WN5   & W   &  E   & E   \\
		WR 7   & WN4   & W   &  W/E & W/E \\
		WR 8   &WN7/WC4&     &      & E   \\
		WR 11  & WC8+0 &     &  E   &     \\              
		WR 16  & WN8h  &     &  W/E & W/E \\              
		WR 18  & WN4   & W   &  W/E & W/E?\\              
		WR 30  & WC6+O &     &  W/E &     \\              
		WR 31  & WN4+O &     &  E   &     \\              
		WR 38  & WC4   &     &  E?  &     \\              
		WR 40  & WN8h  & W/E &  W/E & W/E \\              
		WR 42  & WC7+O &     &  E   &     \\              
		WR 42d & WN4   &     &  W/E &     \\              
		WR 52  & WC5   & R   &  E/R &     \\              
		WR 54  & WN4   &     &  E/R &     \\              
		WR 57  & WC7   &     &  E?  &     \\              
		WR 60  & WC8   &     &  E?  &     \\              
		WR 68  & WC7   &     &  W/E &     \\              
		WR 71  & WN6+? &     &  E   & E   \\              
		WR 75  & WN6   & W/E &  W   & W/E \\              
		WR 85  & WN6   & R   &  W/E &     \\              
		WR 91  & WN7   &     &  W/E & R/E?\\              
		WR 94  & WN6   &     &  E   &     \\              
		WR 98  &WC7/WN6&     &  W/E &     \\
		WR 101 & WC8   &     &  W/E &     \\
		WR 102 & WO2   & W   &  W   & R/E \\              
		%WR 113 & WC8   &     &      &     \\              
		WR 116 & WN8h  &     &      & E   \\              
		WR 124 & WN8h  & E   &      & E   \\              
		%WR 128 & WN4   &     &      &     \\              
		WR 131 & WN7   & R   &      &     \\              
		%WR 132 & WC6   &     &      &     \\              
		%WR 133 & WN4+O &     &      &     \\              
		WR 134 & WN6   & W   &      &     \\              
		WR 136 & WN6h  & W/E &      & W/E \\              
		%WR 153 & WN6   &     &      &     \\              
		%\citet{2010AJ....139.2330W} -52 & WN7 & & & W/E? \\
		\hline                                          
	\end{tabular}                                    
	                                                 
	\medskip
	
	\textit{a}: from \citet{H00}, except WR 8 from \citet{C95}\\
	\textit{b}: uses the categories of \citet{1991IAUS..143..349C}
\end{table}

\begin{figure*}
	\centering 
	\includegraphics[width=170mm]{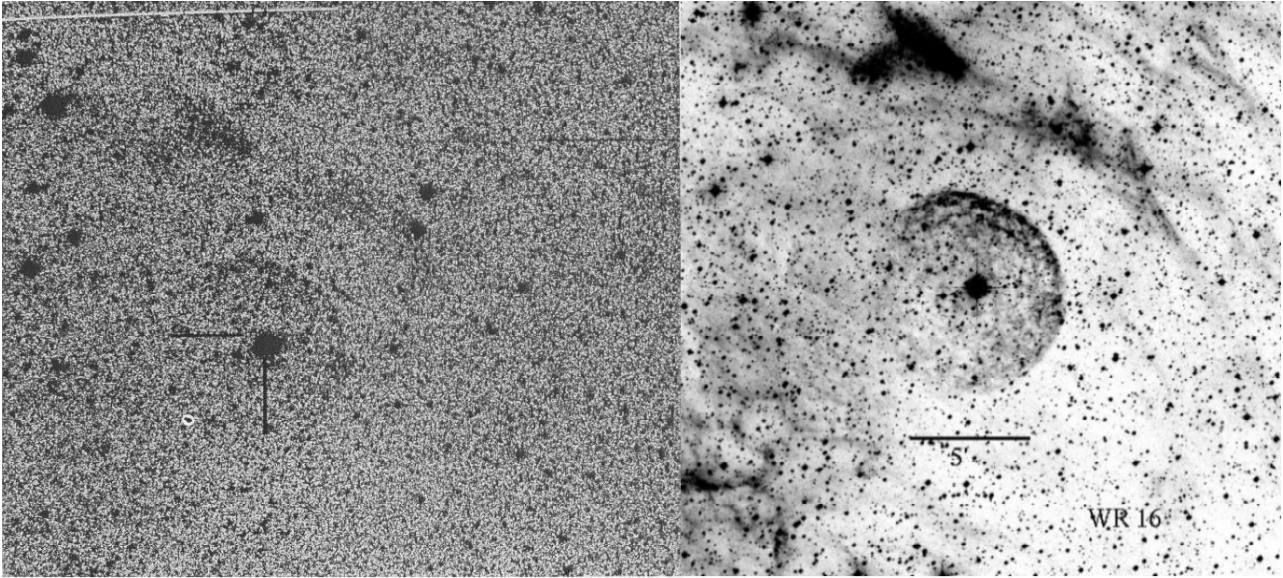}
	\caption{WR 16 (WN8h) The image from \citet{M97} is shown on the left, the image from the SHS is shown on the right. The main orb of nebulosity around WR16 can be clearly seen in the SHS image while it is less prominent in the left hand image. In the SHS image a second and possibly a third concentric ring section can also be discerned. In these images north is up and east is to the left.  }
	\label{WR16BOTH}
\end{figure*}

The survey of \citet{M97} listed 22 southern Galactic WR ejecta nebulae, as opposed to the 10 that we find in the SHS imagery of the same region. There are several reasons for this disparity. We omitted WR stars that were in clusters or were heavily embedded within H~{\sc ii} regions, this accounts for 8 of the \citet{M97} sample which are classified as having an E component (WR 6, 11, 18, 38, 42, 68, 85 and 98) which will not be discussed in section \ref{comments}.

Secondly imagery used by \citet{M97} was obtained using the Curtis Schmidt 0.6m telescope at the Cerro Tololo Inter-American Observatory (CTIO) in Chile, with a resolution of 1.94 arcseconds per pixel compared to the SHS resolution of 0.66 arcseconds per pixel. A direct comparison of SHS and \citet{M97} images of the clear nebulosity around WR 16 is shown in Figure \ref{WR16BOTH}.

%%%%%%%%%%%%INDIVIDUAL OBJECTS%%%%%%%%%%%%%%%%%%%%%%%%%%%%%%%%%%%%%%%
%%%%%%%%%%%%%%%%%%%%%%%%%%%%%%%%%%%%%%%%%%%%%%%%%%%%%%%%%%%%%%%%%%%%%

\subsection{Notes on Individual Objects}
\label{comments}

\subsubsection*{WR 8 - HD 62910}

\begin{figure}
	\centering
	\includegraphics[width=80mm]{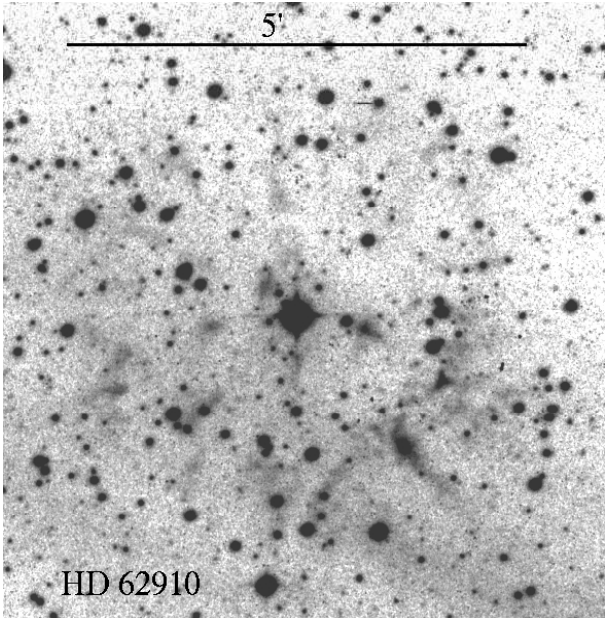}
	\caption{The SHS H$\alpha$ image of the field around HD 62910 (WR 8,  WCE+) shows newly revealed nebulosity. North is up, east is to the left.}
	\label{WR8}
\end{figure}

In the SHS imagery of the region around WR 8 (Figure \ref{WR8}) one can clearly discern nebulosity that appears to be associated with the star. It is aligned along radial ``spokes" with an especially prominent example to the south-west. These spokes define a circular structure approximately 5 arcminutes in diameter, internal to which there are several prominent clumps of nebulosity. 

The spectral type of the host star is listed by \citet{H00} as WN7/WC4. Its spectrum places WR 8 neatly between those of WN and WC stars \citep{C95}. This was initially interpreted as a sign of binarity - a system comprising both WC and WN stars - however \citet{C95} showed that the wind properties were the same for both the N and C components - implying a single star origin. 

The presence of an ejecta nebula around such an unusual host star - a star possibly seen during the transition from the WN to the WC phase - is especially interesting since the composition of the nebula could be helpful towards understanding the evolutionary state of the star.

\subsubsection*{WR 30 - HD 94305}

\begin{figure}
	\centering
	\includegraphics[width=80mm]{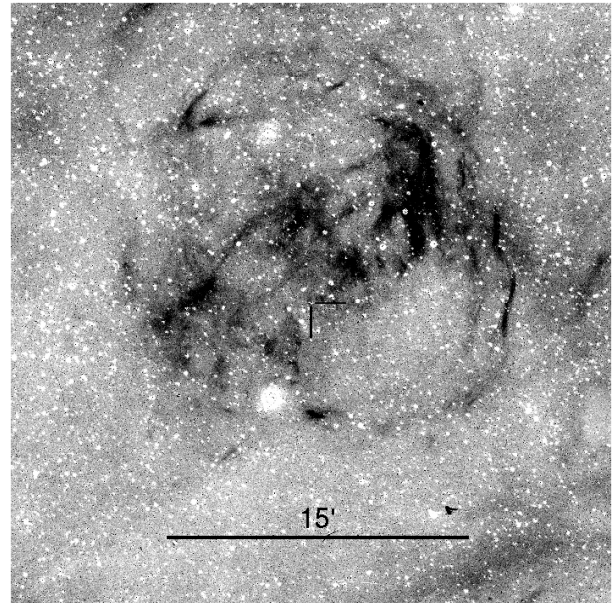}
	\caption{SHS image subtraction (H$\alpha$ - Short Red) of the field around WR 30 (WC6+O). The filamentary structures were listed as a ring nebula of W/E type by \citet{M97}. North is up and east is to the left. }
	\label{WR30}
\end{figure}

WR 30 was listed as posessing a ring nebula by \citet{M97}. The SHS imagery shows some filamentary structures around the star, some of which take the form of arcs roughly centred on the star, shown in \ref{WR30}. The structure shown appears to be related to the star however there is no morphological evidence in the form of flocculence within the arcs to indicate that this nebulosity contains ejecta. 

\subsubsection*{WR 31 - HD 94546}

\begin{figure}
	\centering
	\includegraphics[width=80mm]{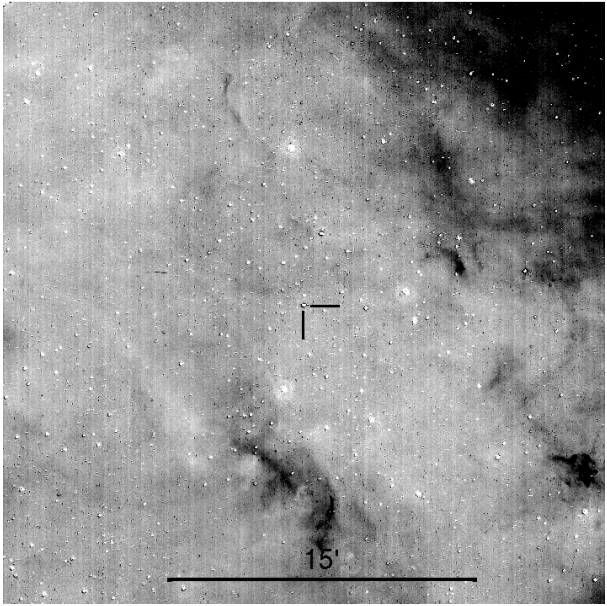}
	\caption{SHS image subtraction (H$\alpha$ - Short Red) of the field around WR 31 (WN4+O7). The very faint, indistinct arc of emission to the west was suggested to be part of a ring by \citet{M94b} and \citet{M97}. North is up, east is to the left. }
	\label{WR31}
\end{figure}

The nebula around this object, shown in Figure \ref{WR31}, was classified by \citet{M94b} as an R type ring nebula, with a diameter of 6.7 arcminutes and was later upgraded to ejecta (E) type by \citet{M97}. One can perhaps see faint traces of what appears to be an arc to the east of the star, but this is inconclusive and there is no other evidence for structure on the scales indicated by Marston or that this could be an E type WR nebula.

\subsubsection*{WR 52 - HD 115473}

\begin{figure}
	\centering
	\includegraphics[width=80mm]{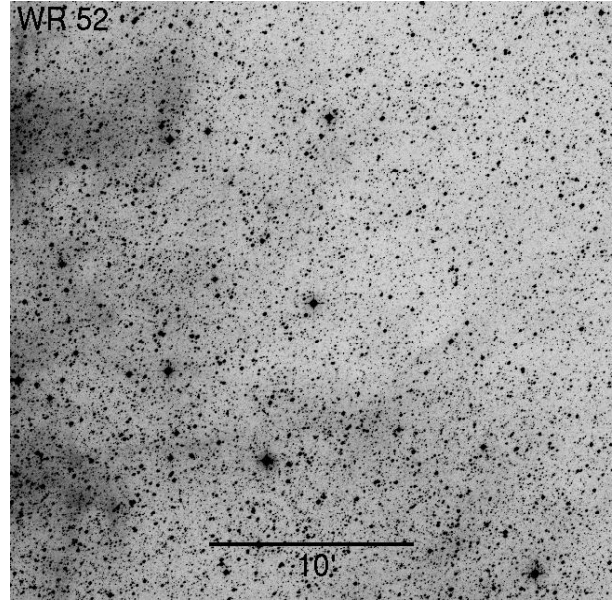}
	\caption{SHS image of the region around WR 52 (WC5). The faint diffuse emission has been suggested to be part of a ring, however this is not supported by the SHS imagery. North is up, east is to the left. }
	\label{WR52}
\end{figure}

The nebulosity around this object, shown in Figure \ref{WR52}, was identified by \citet{M94a} as a possible ring nebula with radius 60 arcminutes. Later it was classified as an R type nebula by \citet{M97} because of diffuse [O~{\sc iii}] emission. The region which Marston identified as being one half of a ring is shown in the SHS imagery but does not appear to exhibit any ring-like structures. Although WR 52 appears to be embedded in diffuse H$\alpha$ emission, there appears to be no evidence for an ejecta nebula.

\subsubsection*{WR 54}

\begin{figure}
	\centering
	\includegraphics[width=80mm]{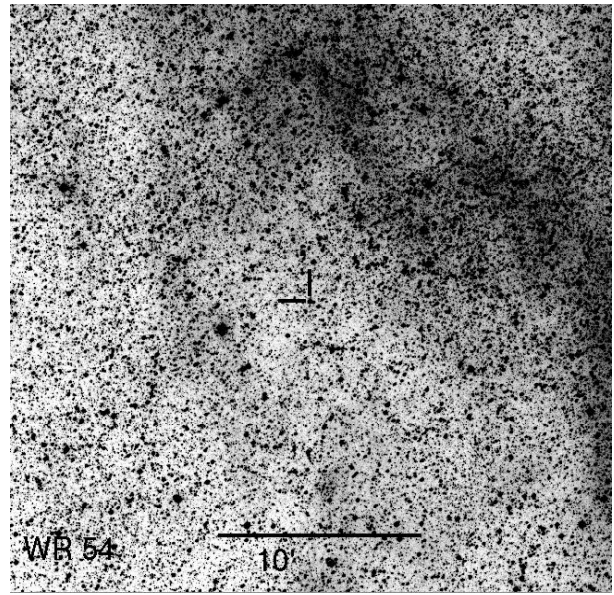}
	\caption{SHS image of the region around WR 54 (WN4). The filamentary nebulosity surrounding the star has been proposed to possibly be part of a ring. North is up, east is to the left.}
	\label{WR54}
\end{figure}

This object, shown in Figure \ref{WR54}, was noted as a possible faint ring nebula of radius 20 arcminutes by \citet{M94a}. Subsequently this object was classified as an E/R type by \citet{M97}. In the SHS image we can see an apparent arc of nebulosity to the NW but this appears to be filamentary and of dubious relation to the star.  

\subsubsection*{WR 57}

\begin{figure}
	\centering
	\includegraphics[width=80mm]{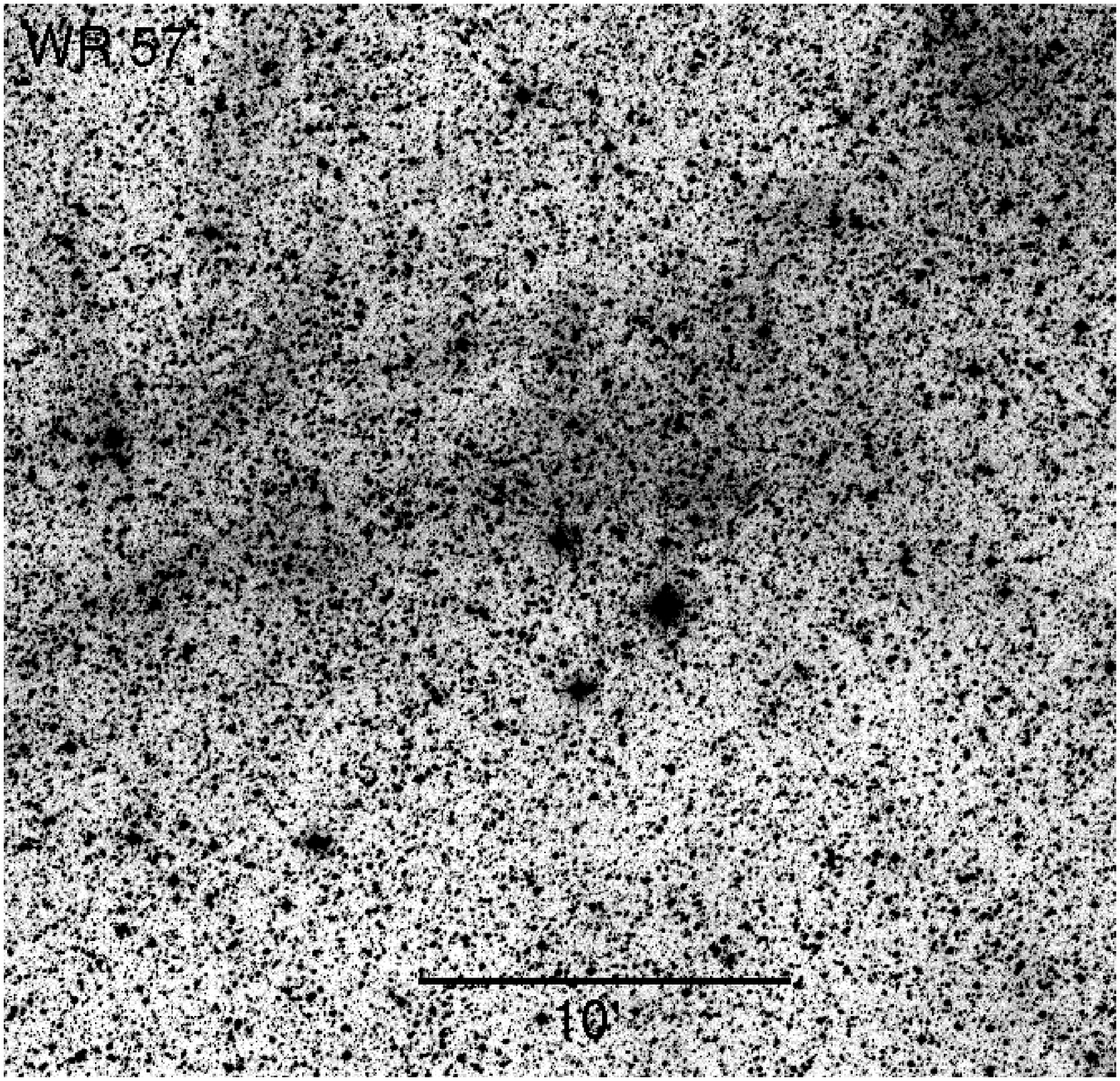}
	\caption{SHS image of the region around WR 57 (WC7). The filament running east-west in this image was suggested by \citet{M97} to be a possible ring section. North is up, east is to the left.}
	\label{WR57}
\end{figure}

WR 57 was suggested to have a possible ejecta nebula of radius 8 arcminutes by \citet{M97}. The SHS image, Figure \ref{WR57}, shows a filament running E-W across the image, with no obvious evidence of a connection with WR 57.

\subsubsection*{WR 60}

\begin{figure}
	\centering
	\includegraphics[width=80mm]{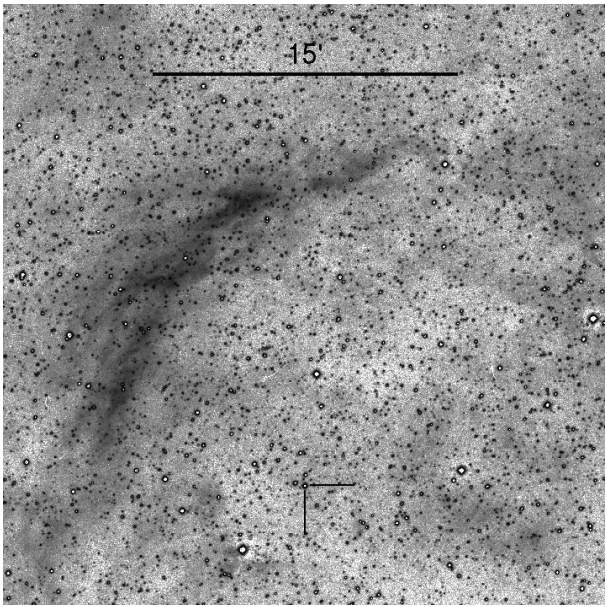}
	\caption{SHS image subtraction (H$\alpha$ - Short Red) of the field around WR 60 (WC8). This provides confirmation of the ring section previously detected by \citet{M94b}. North is up, east is to the left.}
	\label{WR60}
\end{figure}

WR 60 was claimed to have a ``90\% complete" ring nebula with a radius of 9 arcminutes by \citet{M94b} and later tentatively ascribed E type status by \citet{M97}. The SHS image, shown in Figure \ref{WR60}, confirms the detection of a possible ring section to the NE of WR 60. However there is no evidence of the star being further encircled beyond this, suggesting that this nebulosity is merely a diffuse filament. 

\citet{S09} show that the arc of nebulosity possesses spectral features similar to those associated with supernova remnants (SNRs) and propose a new designation for this nebula as an SNR - G310.5 + 0.8.  

\subsubsection*{WR 71}

\begin{figure*}
	\centering
	\includegraphics[width=170mm]{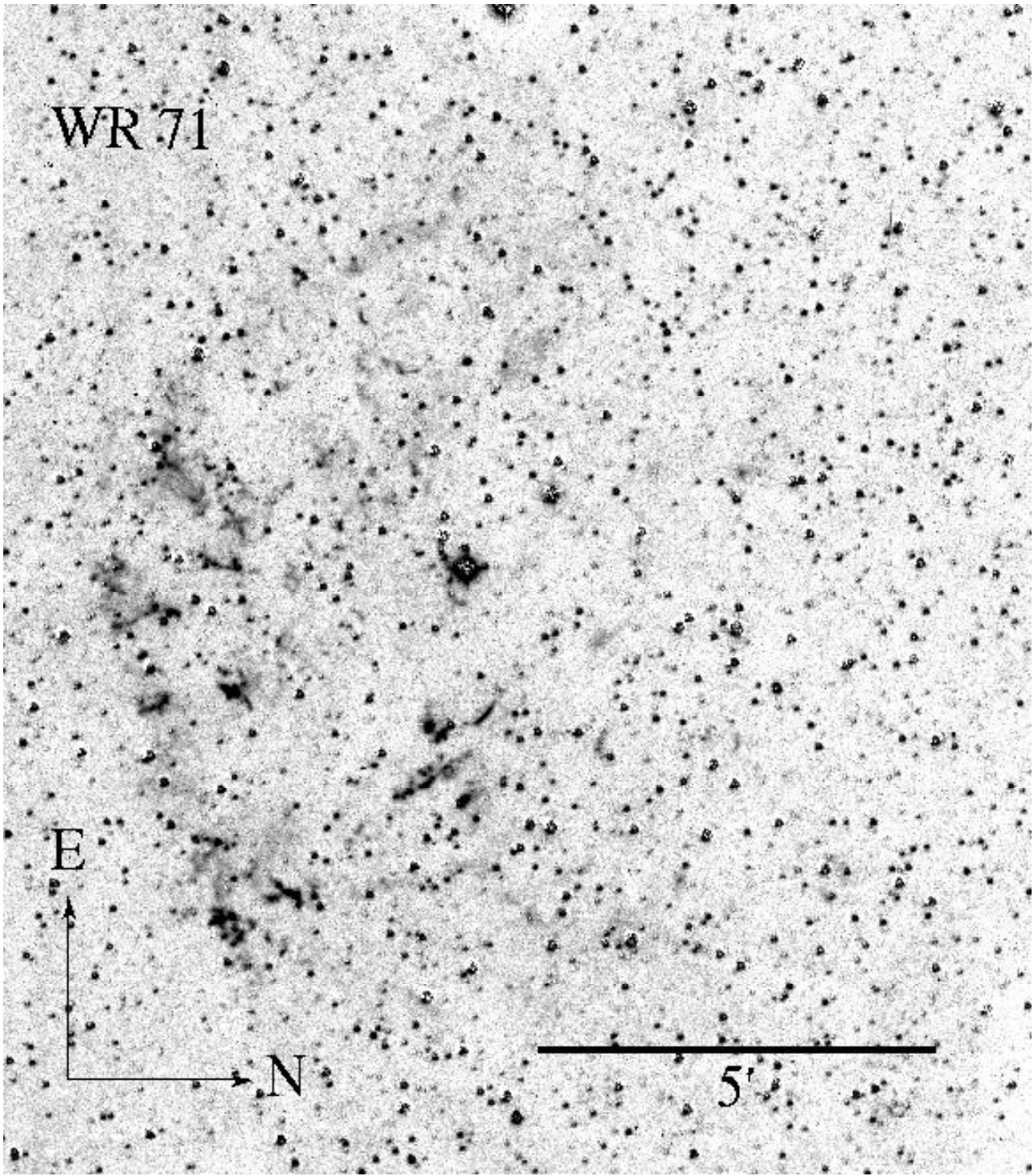}
	\caption{SHS image  of  the field around WR 71 (WN6) (H$\alpha$-red subtraction). The tenuous flocculent nebulosity to the south strongly suggests stellar ejecta, as noted by \citet{M94b}. }
	\label{WR71}
\end{figure*}

The SHS image, shown in Figure \ref{WR71}, confirms the tenuous nebula around WR 71 first noted by \citet{M94b} and later classified as an E type nebula by \citet{M97}. It appears similar to RCW 58 and anon (WR 8), in that it has highly clumped nebulosity to the south, although much fainter than either of the above. The SHS subtracted (H$\alpha$-R) image (Figure \ref{WR71}) also shows some arcs to the west while improving on the detail of the flocculent structure to the south. The progenitor is a runaway star and as such is significantly out of the plane ($z = 1190$ pc) suggesting that this object may have unique kinematics due to the low ISM density at this  z distance. For further discussion of this object see Section \ref{sec:bin}.

\subsubsection*{WR 91}

\begin{figure}
	\centering
	\includegraphics[width=80mm]{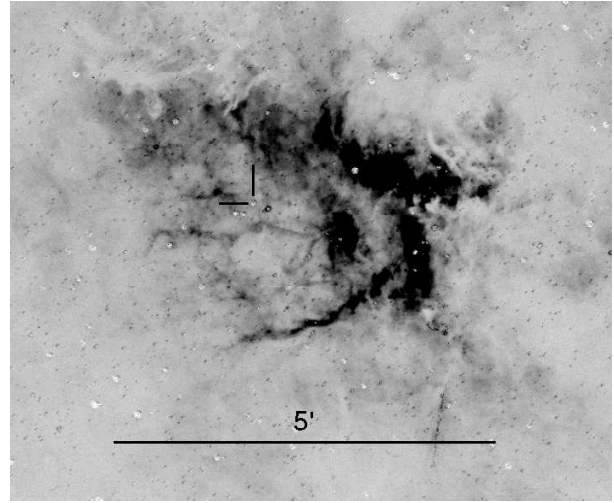}
	\caption{SHS image subtraction (H$\alpha$ - Short Red) of the field around WR 91 (WN7). North is up, east is to the left.}
	\label{WR91}
\end{figure}

For the region around WR 91, shown in Figure \ref{WR91}, the nebulosity was classified as W/E by \citet{M97}. The structure, however, is not centred on WR 91 in the manner that would be expected of a nebulosity created by a stellar wind. Instead the associated diffuse nebulosity appears to be radiatively excited.

\subsubsection*{WR 94}

\begin{figure}
	\centering
	\includegraphics[width=80mm]{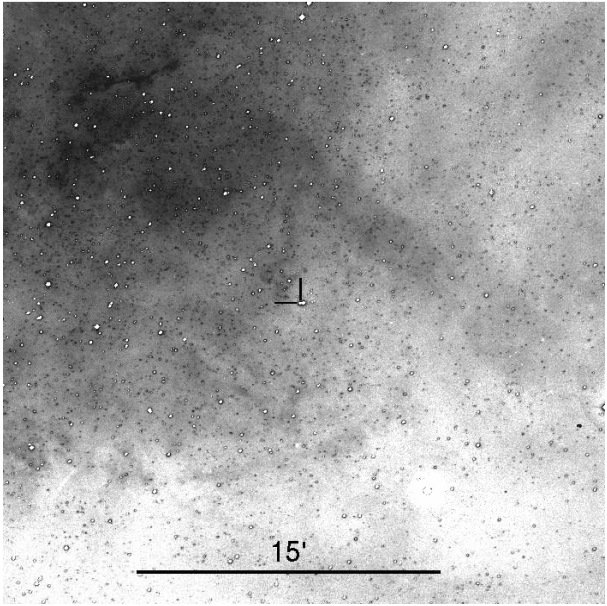}
	\caption{SHS image subtraction (H$\alpha$ - Short Red) of the field around WR 94 (WN6). North is up, east is to the left.}
	\label{WR94}
\end{figure}

The nebulosity around WR 94, shown in Figure \ref{WR94}, appears to be radiatively excited by the star. It displays no signs of clumpiness or flocculence and hence we are unable to confirm this as an ejecta nebula of radius 11 arcminutes as suggested by \citet{M97}.

\subsubsection*{WR 101}

\begin{figure}
	\centering
	\includegraphics[width=80mm]{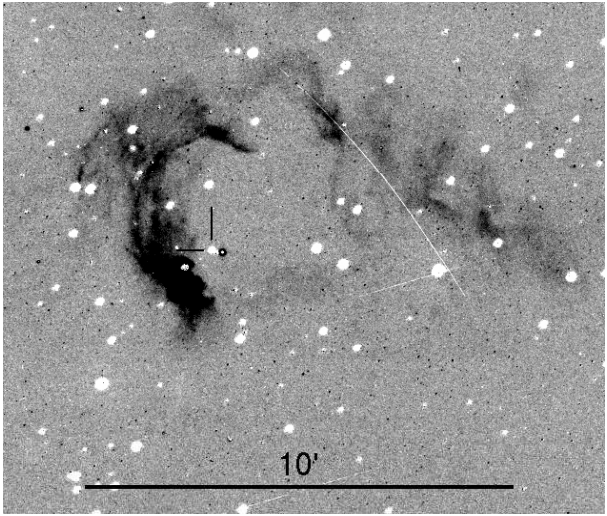}
	\caption{SHS image subtraction (H$\alpha$ - Short Red) for the field around WR 101 (WC8). North is up, east is to the left.}
	\label{WR101}
\end{figure}

The strangely shaped nebula near WR 101, shown in Figure \ref{WR101}, was detected and classified as a W/E type nebula by \citet{M94b}. Upon inspection of the SHS imagery it is clear that there is a main clump to the south east of the star that is reminiscent of other ejecta type nebulae, however it is unusual in that it appears to be so highly asymmetric, resembling the northern clump in the Bubble Nebula (NGC 7635) around the Of-type star BD+602522. The arcs of nebulosity to the north and east are possibly associated with the star. \citet{CGP02} showed that the ionized mass of the nebulosity is of the order $200 M_{\sun}$ for a distance of 3.2 kpc, too high to be pure ejecta. Thus this is likely a swept up ISM nebula with at best some processed component.

\subsubsection*{A48: \citet{2010AJ....139.2330W} - 50}

\begin{figure}
	\centering
	\includegraphics[width=80mm]{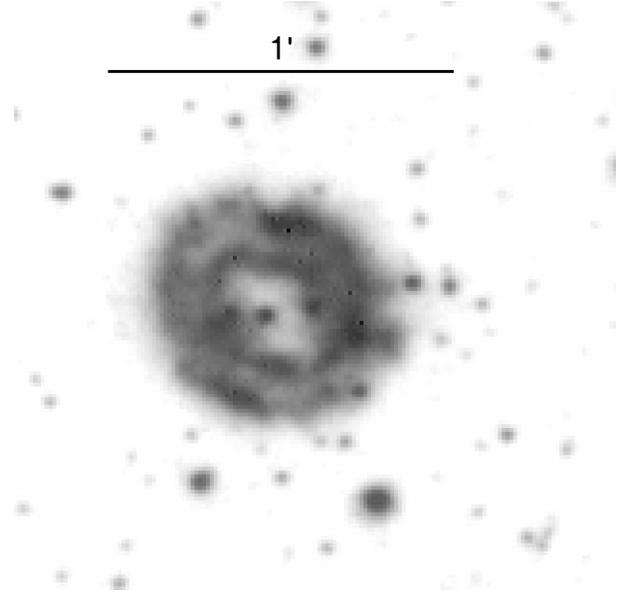}
	\caption{SHS image for the field around A48 (WN6). North is up, east is to the left.}
	\label{W52}
\end{figure}

This object, discovered by \citet{1966ApJ...144..259A} appears to have two very distinct complete rings.  The possible implications of nebular detections with multiple rings are discussed in Section \ref{Mrings}. The \citet{2010AJ....139.2330W} ring seen at 8 and 24 $\mu$m is coincident with the optical ring which appears in the R band as well as H$\alpha$.

The SHS image lacks the resolution to determine whether the rings have any flocculent structure and as such it is difficult to classify this nebulosity beyond the default classification of W type. The Spitzer 8$\mu$m image of A48 presented by \citet{2010AJ....139.2330W} appear to have higher resolution and show more structure in the inner ring. This implies that the inner ring is likely to be composed of ejected stellar material, if this is the case then the classification will be W/E.

\subsubsection*{BAT99-11 - HD 32402}

\begin{figure}
	\centering
	\includegraphics[width=80mm]{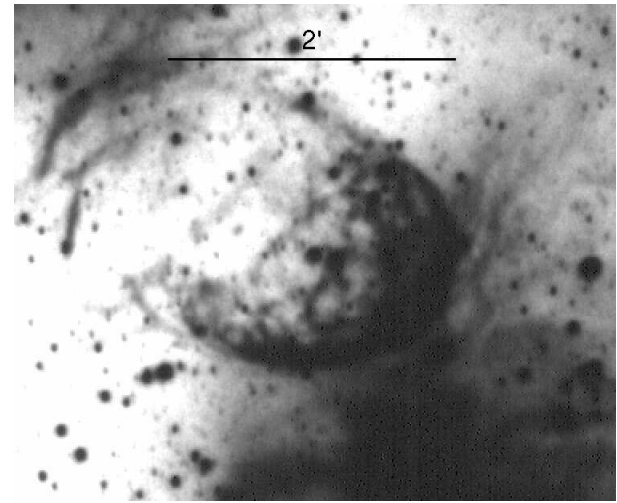}
	\caption{SHS image for the field around the LMC star HD 32402 (WC4). North is up, east is to the left.}
	\label{BAT11}
\end{figure}

The nebulosity around BAT99-11 appears ovoid in the same manner as NGC 6888, a bright shock with clear internal structure. If a bona-fide WR nebula it is the largest with the main bubble having a radius of almost 13 pc and distinct arcs at still greater distances. In the SHS imagery it is impossible to make out the internal structure beyond several filaments confined to the interior of the main shock.

We classify this object as a W/E type nebula based on the internal structure and obvious crescent ring section created by wind interactions.

\subsubsection*{BAT99-15 - HD 268847}

\begin{figure}
	\centering
	\includegraphics[width=80mm]{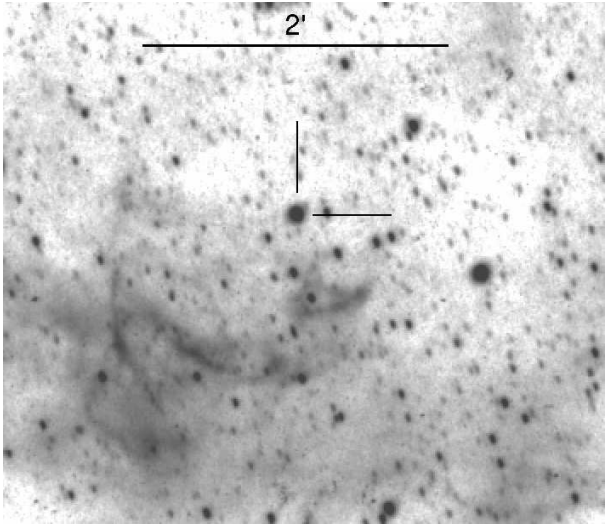}
	\caption{SHS image for the field around the LMC star HD 268847 (WN4b). North is up, east is to the left.}
	\label{BAT15}
\end{figure}

BAT99-15 has two clear nebulous arcs to the south. These arcs reside in a region of diffuse H$\alpha$ emission suggesting that they are the interaction between stellar wind and this diffuse material. Given the faintness of this object however it is impossible to clearly rule in or out any E type contribution to this W type nebulosity.

\subsubsection*{BAT99-16 - HD 33133}

\begin{figure}
	\centering
	\includegraphics[width=80mm]{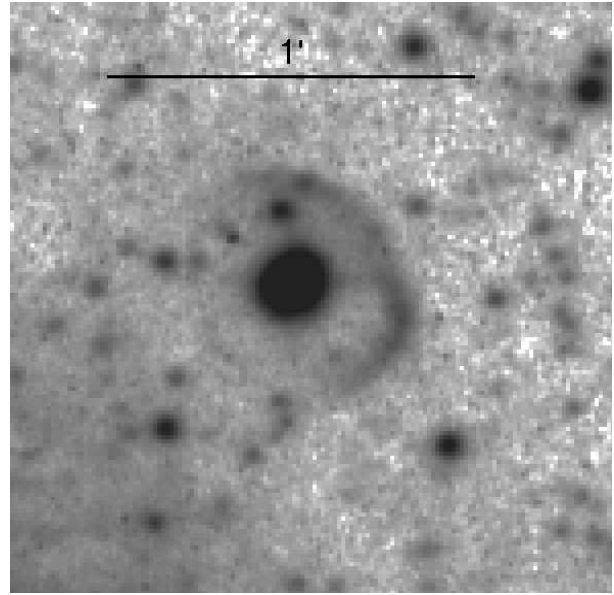}
	\caption{SHS image for the field around the LMC star HD 33133 (WN8h). North is up, east is to the left.}
	\label{BAT16}
\end{figure}

The crescent nebulosity to the north and west of BAT99-16 is deceptively small, the arc of radius 3.8 pc is actually larger than for NGC 6888 which has a radius of 3.0 pc. As with BAT99-15 we cannot see any internal structure, so it is impossible to ascertain whether the classification should be W or W/E.

\subsubsection*{BAT99-65}

\begin{figure}
	\centering
	\includegraphics[width=80mm]{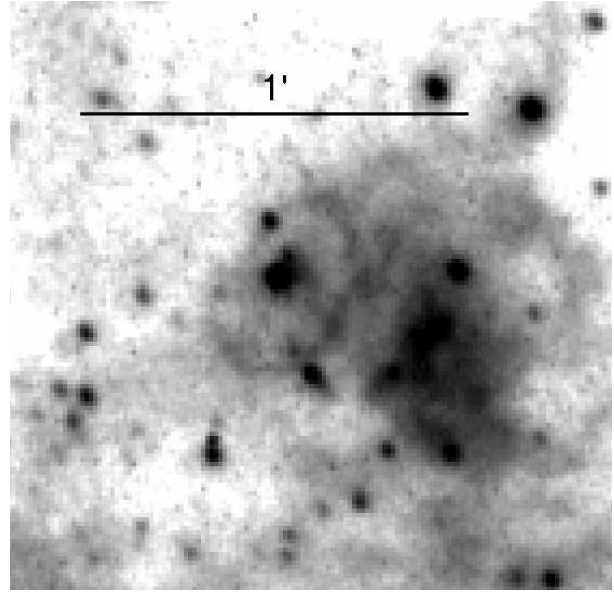}
	\caption{SHS image for the LMC field around BAT99-65 (WN4). North is up, east is to the left.}
	\label{BAT65}
\end{figure}

The oval ring nebula around BAT99-65 suggests a W type origin. Again, we cannot reliably ascribe any further classification to this object.

%%%%%%%%%%%%%%%%%%%%%%%%%%%%%%%%%%%%%%%%%%%%
%%%%%%%%%%%%%%%%%%%%%%%%%%%%%%%%%%%%%%%%%%%%

\section{Discussion}

In Table \ref{ProgTable} we summarise the properties of each nebula detected in our MC and LMC survey along with the implied physical extent. The fractions of WR nebulae in the MW or LMC with either WN or WC central stars are presented in Table \ref{StatsTable} along with the fractions for the complete WR population.

\begin{table*}
	\centering
	\caption{Wolf-Rayet stars with Ejecta Nebulae}
	\label{ProgTable}
	\begin{tabular}{lp{1.5cm}llp{1.5cm}p{1.5cm}p{1.5cm}l}
		\hline
		Nebular Name & Stellar Catalogue Number$^a$ & Name & Spectral Type$^a$ & Angular Radius (arcmin) & Distance$^b$ (kpc) & Radius (pc) & $\left|z\right|^a$ (pc) \\
		\hline
		Milky Way \\
		\hline
		NGC 2359  & WR 7   & HD 56925       & WN4     & 2.2  & 3.7 & 2.3  & 8   \\
		Anon      & WR 8   & HD 62910       & WN7/WC4 & 1.9  & 3.5 & 1.9  & 230 \\
		Anon      & WR 16  & HD 86161       & WN8h    & 3.7  & 2.4 & 2.5  & 110 \\
		NGC 3199  & WR 18  & HD 89358       & WN4     & 4.8  & 2.2 & 3.1  & 37  \\
		RCW 58    & WR 40  & HD 96548       & WN8h    & 3.3  & 2.3 & 2.2  & 190 \\
		Anon      & WR 68  & Anon           & WC7     & 6.6  & 3.3 & 6.3  & 110 \\
		Anon      & WR 71  & HD 143414      & WN6+?   & 4.5  & 9.0 & 12 & 1200\\
		RCW 104   & WR 75  & HD 147419      & WN6     & 9.3  & 2.2 & 5.9  & 56  \\
		Anon      & WR 91  & Anon           & WN7     & 3.7  & 7.2 & 7.7  & 130 \\
		G 2.4+1.4 & WR 102 & Anon           & WO2     & 3.1  & 5.6 & 5.0  & 140 \\
		Anon      & WR 116 & Anon           & WN8h    & 3.3  & 2.5 & 2.4  & 14  \\
		M 1-67    & WR 124 & Merrill's Star & WN8h    & 0.7  & 3.4 & 0.70 & 190 \\
		NGC 6888  & WR 136 & HD 192163      & WN6h    & 8.2  & 1.3 & 3.0  & 53  \\
		A48$^c$   &        &                & WN6     & 0.3  &     &      &    \\
		
		\hline
		LMC Candidates  \\
		\hline
		Anon      & BAT99-11 & HD 32402  & WC4     & 0.88 & 50 & 12.8 \\
		Anon      & BAT99-15 & HD 268847 & WN4b    & 0.66 & 50 & 9.6  \\
		Anon      & BAT99-16 & HD 33133  & WN8h    & 0.26 & 50 & 3.8  \\
		%Anon      & BAT99-38 & HD 36402  & WC4+O8I & 0.21 & 50 & 3.0  \\
		%Anon      & BAT99-62 & Anon      & WN3o    & 0.18 & 50 & 2.5  \\
		Anon      & BAT99-65 & Anon      & WN4     & 0.23 & 50 & 3.4  \\
		\hline
	\end{tabular}
	
	\medskip
	
	\textit{a}: from \citet{H00} (Galactic).\\
	\textit{b}: Galactic distances from \citet{H00}, LMC distance \citet{M06}\\
	\textit{c}: \citet{2010AJ....139.2330W} - 50

\end{table*}

\begin{table}
	\centering
	\caption{Results of Survey}
	\label{StatsTable}
	\begin{tabular}{p{3cm}rrp{1.5cm}r}
		\hline
		                            & WN   & WC    & (WN/WC \& WO) & WR  \\
		\hline
		\bf{Milky Way$^a$} \\
		WR stars with Ejecta nebula & 10   & 1     & 2    & 13   \\
		Total WR stars              & 127  & 87    & 13   & 227  \\
		Ratio                       & 0.08 & 0.01  & 0.15 & 0.06 \\
		\hline
		\bf{LMC$^b$} \\
     WR stars with candidate Ejecta nebulae & 3    & 1     &  0    & 4    \\
		Total WR stars              & 108  & 24    &  2    & 134  \\
		Ratio                       & 0.03 & 0.04  &  0.00 & 0.03 \\
		\hline
	\end{tabular}
	
	\medskip
	\textit{a}: Not including WR stars discovered by \citet{2010AJ....139.2330W} as they were found by examining central stars of detected IR ring nebulae.\\
	\textit{b}: Counting all O3If*/WN6-A stars as isolated WN type.\\
	%\textit{d}: Assuming each of our candidate nebulae contains ejecta.\\	

\end{table}

\subsection{Multiple Rings?}\label{Mrings}

Several WR stars have been suggested to have multiple rings, e.g. two by \citet{M95}, however WR 16 represents the clearest example of a multiple ringed object as both concentric rings are clearly visible in H$\alpha$ (Figure \ref{WR16BOTH}). The inner ring is undisputably created by stellar outflows, as has been confirmed spectroscopically \citep{M99} - the outer ring though could be swept up material. If the outer ring contains some component, however small, of processed material, the relative compositions and kinematics of the structures could provide useful constraints on the evolution of the star. 

The recent discovery of a new double ring structure surrounding a WR star (\citet{2010AJ....139.2330W} - 50) presents another opportunity to study multiple epochs of a star's history. The SHS imagery does not reveal enough detail to reveal whether the inner ring is likely to be ejected material. However, as mentioned in Section \ref{comments}, there is some evidence of knots in the inner ring in the 8 $\mu$m image presented by \citet{2010AJ....139.2330W}. 

\subsection{Central Star Properties}

The most prominent WR ejecta nebulae are associated with WNh stars, namely: WR 16, 40, 124 and 136. These nebulae have all been confirmed to contain nucleosynthetic products. WNh stars still have some hydrogen left in their atmospheres, i.e. the loss of the H dominated envelope is not complete \citep{C07}. The fact that there are bright, N and/or He enriched nebulae around such objects is consistent with the material formerly having made up part of their envelopes. That the shells of matter around these stars are broadly circular indicates that the beginning of the WR phase for these objects may have been dominated by an extreme mass loss event which created the nebulae.

In section \ref{comments} we commented on the unusual spectral type of WR 8, (WN7/WC4); \citep{C95}. This transition star's nebulosity is strongly reminiscent of RCW 58 - a confirmed ejecta nebula. Assuming that it is an ejecta nebula, the nebula around WR 8 may provide an opportunity to spectroscopically probe matter ejected prior to or during this transition phase.

\begin{table}
	\centering
	\caption{WR Nebulae with Binary Central Stars}
	\label{BinaryTable}
	\begin{tabular}{p{3cm}rrr}
		\hline
		             & Isolated WR & Binary WR & All WR    \\
		\hline
		\bf{Milky Way$^a$} \\
		Ejecta Nebulae & 12   & 1    & 13  \\
		All WR Stars   & 141  & 86   & 227 \\
		Ratio          & 0.09 & 0.01 & 0.06\\
		\hline
		\bf{LMC$^b$} \\
      Candidate Ejecta Nebulae & 4     & 0    & 4   \\
		All WR Stars   & 102   & 32   & 134 \\
		Ratio          & 0.04  & 0.00 & 0.03\\
		\hline
	\end{tabular}

	\medskip
	\textit{a}: Not including WR stars discovered by \citet{2010AJ....139.2330W}.\\
	\textit{b}: Counting all O3If*/WN6-A stars as isolated WN type.
	
\end{table}

\subsection{Binarity} \label{sec:bin}

From Table \ref{ProgTable} a curious fact emerges: only two out of the nineteen WR Ejecta nebula central stars listed here are binaries. In Table \ref{BinaryTable} we summarise the binary fractions for the WR populations in the MW and LMC along with the fraction that we have identified as ejecta nebulae for each case. The binarity classifications of \citet{H00} and \citet{B99} were adopted for the Galactic and LMC populations respectively. It is striking that the fraction of binary WR stars having ejecta nebulae is so low.

If the fraction of WR stars with ejecta nebulae was the same for binaries as single WR stars we would expect $\sim 10\%$ of WR binary stars to possess ejecta nebulae - which translates to around 6 expected in the MW compared to 1 observed (WR 71). In the LMC there is one binary WR star with what appears to be an ejecta type nebula, whereas we might expect $\sim$ 4 LMC binary WR stars to possess ejecta nebulae.

It has long been speculated that there are two methods of creating WR stars, mass transfer between binary partners and mass loss of an isolated star (e.g.  \citealt{SP68}). A possible reason for the relative absence of ejecta nebulae around binary WR stars is that mass transfer onto a companion inhibits the mechanism that produces an ejecta nebulae around single WR stars. 

The nebulosity surrounding WR 71 (see Figure \ref{WR71}) is an exception to the previous discussion. The progenitor star is of spectral type WN6+? \citep{H00}. \citet{I83} suggested that the binary partner is a low mass evolved stellar remnant, either a neutron star or a black hole and that the supernova which created the collapsed object likely occured when the system was in the disk of the MW. The loss of mass which occured then left the binary companion (which we now see as WR 71) travelling along whichever velocity vector it possessed at the moment of the supernova. The large elevation of WR 71 above the Galactic plane ($z=-1185$ pc) suggests that a significant component of this velocity was in the $z$ direction. The nebulosity is physically much larger than counterparts in the plane because the density of the ISM at this elevation is much lower, implying that all the circumstellar material is ejecta, as there is so little ISM to be swept up. 

It is odd, considering the previously noted anticorrelation of ejecta nebulae and binary WR systems, that such a clear ejecta nebulae should surround such a star. However the explanation may lie in the very low mass of the unseen binary companion ($\sim 3M_{\sun}$ - \citealt{I83}), since normally WR binary companions are O stars of $\sim 30 M_{\sun}$ (\citet{H00} - Tables 18-19). A low mass binary companion may not influence the later creation of an ejecta nebula to the same degree as a less evolved high mass companion. This echoes the work of \citet{NF94}, who discussed different scenarios for massive binary star evolution. If the WR 71 nebulosity originated in this way, then the material that comprises the nebula would be expected to have the same enrichment pattern as other WR nebulae. If, on the other hand, the nebulosity is the product of more complex binary interaction and mass transfer then its composition would be more difficult to predict.

This suggests that over the history of a binary system with two close stars of high initial mass the more massive partner will not create an ejecta nebula -  all the mass will be accreted by the partner - while the initially lower mass star can create an ejecta nebula as its mass loss is not as influenced by its low mass post-SN companion.

\section{Summary}

The morphological classification scheme of \citet{C81,1991IAUS..143..349C} has been discussed and compared with spectroscopically derived abundances of WR nebulae and a modified set of criteria for ejecta nebulae around WR stars derived.

Using SHS H$\alpha$ survey data we have examined the environs of each WR star in the Milky Way and Magellanic Clouds for evidence of the presence of  nebulae that could be composed of stellar ejecta.  This has yielded one new strong candidate (WR 8), confirmed the morphology of another (WR 71) and shown several previously claimed examples to be unlikely candidates.
 
A prevalence of WNh subtypes amongst ejecta nebulae central stars was found. In addition an anti-correlation between WR binarity and the occurence of an associated nebulosity was also found. It is speculated that this may be because binary interactions and mass transfer may inhibit the formation of an ejecta nebula. This hypothesis is discussed in light of the status of WR 71 - a binary runaway star with an ejecta nebula.

\section*{Acknowledgments}

The LMC section of this work would not have been possible without the contribution of Mike Read from the Wide Field Astronomy Unit at the University of Edinburgh and his efforts to `rescue' the Magellanic Clouds portion of the SHS. 

We would also like to thank Dr A. P. Marston for his comments on an early draft of this paper.

Thanks also to the RAS library for providing physical access to copies of ApJSS and allowing the author to scan relevant plates from \citet{M97} at higher quality than those provided in the online versions. 

This research made use of Montage, funded by the National Aeronautics and Space Administration's Earth Science Technology Office, Computation Technologies Project, under Cooperative Agreement Number NCC5-626 between NASA and the California Institute of Technology. Montage is maintained by the NASA/IPAC Infrared Science Archive.

D.J.S. is supported fully by an STFC postgraduate studentship.

%\bibliographystyle{mn2e}
%\bibliography{WRpaper}

\label{lastpage}

\end{document}